\documentclass[prb,showpacs]{revtex4}
\usepackage{epsfig}
\usepackage{graphicx}
\usepackage{amsmath}
\usepackage{bm}
\usepackage{amssymb}
\usepackage{color}

\newif\ifold             \oldtrue            
\def\ba{\begin{eqnarray}}
\def\ea{\end{eqnarray}}

\newcommand{\be}{\begin{equation}}
\newcommand{\ee}{\end{equation}}

\begin{document}


\title{Dynamics and phase diagram of the $\nu=0$ quantum Hall state in bilayer
graphene}
\date{\today}


\author{E. V. Gorbar}
\affiliation{Bogolyubov Institute for Theoretical Physics, 03680, Kiev, Ukraine}

\author{V. P. Gusynin}
\affiliation{Bogolyubov Institute for Theoretical Physics, 03680, Kiev, Ukraine}

\author{V. A. Miransky}
\altaffiliation[On leave from ]{Bogolyubov Institute for Theoretical Physics, 03680, Kiev, Ukraine.}
\affiliation{Department of Applied Mathematics, University of Western Ontario, London,
Ontario N6A 5B7, Canada}

\begin{abstract}
Utilizing the Baym-Kadanoff formalism with the polarization function
calculated in the random phase approximation, the dynamics of the $\nu=0$
quantum Hall state in bilayer graphene is analyzed. Two phases with nonzero
energy gap, the ferromagnetic and layer asymmetric ones, are found.
The phase diagram in the plane $(\tilde{\Delta}_0,B)$,
where $\tilde{\Delta}_0$ is a top-bottom gates voltage imbalance, is
described.
It is shown that the energy gaps in these phases
scale linearly,
$\Delta E\sim 10 B[{\mbox T}]{\mbox K}$, with magnetic field.
The comparison of these results with recent experiments
in bilayer graphene is presented.
\end{abstract}

\pacs{81.05.ue, 73.43.-f, 73.43.Cd}


\maketitle
\section{Introduction}
\label{1}

The properties of bilayer graphene
\cite{McC,McC2,Nov,Hen,CN}, consisting of two closely
coupled graphene layers, have attracted great interest.
The possibility of inducing and controlling the energy gap by gates voltage makes
bilayer graphene one of the most active
research areas with very promising  applications in electronic devices. Recent
experiments in bilayer graphene \cite{FMY,Zh} showed the generation of energy gaps
in a magnetic field with complete lifting of the eightfold degeneracy in the
zero energy Landau level, which leads to new quantum Hall states with filling
factors $\nu=0,\pm1,\pm2,\pm3$. Besides that, in suspended bilayer graphene,
Ref. \onlinecite{FMY} reports the observation of an extremely large
magnetoresistance in the $\nu=0$ state due to the energy gap $\Delta E$, which
scales linearly with a magnetic field $B$, $\Delta E \sim 3.5-10.5
B[{\mbox T}]{\mbox K}$, for $B \lesssim 10{\mbox T}$. This linear scaling is
hard to explain by the standard  mechanisms \cite{QHF,MC} of gap generation
used in a monolayer graphene, which lead to large gaps of the order of the
Coulomb energy $e^{2}/l \sim B^{1/2}$, $l=(\hbar c/eB)^{1/2}$ is the magnetic
length.

The theory of the quantum Hall effect (QHE) in bilayer graphene has been
studied in Refs. \onlinecite{Bar,Sh,NCD,NL1,GGM}. In particular, the
gap equation for the quasiparticle propagator including the polarization
screening effects has been recently studied in Refs. \onlinecite{NL1,GGM}. While
a polarization function with no magnetic field  was used in Ref. \onlinecite{NL1},
the polarization function with a magnetic field was utilized in Ref. \onlinecite{GGM}.

In this paper, we study the dynamics of clean bilayer graphene in a magnetic
field, with the emphasis on the $\nu=0$ state in
the quantum Hall effect (QHE) [a brief description of a part of
the results of this study was presented in Ref. \onlinecite{GGM}].
It will be shown that, as in the case of monolayer
graphene \cite{Gorb1}, the dynamics in the QHE in bilayer graphene is described
by the {\em coexisting} quantum Hall ferromagnetism \cite{QHF} (QHF) and
magnetic catalysis \cite{MC} (MC) order parameters. The essence of the dynamics
is an effective reduction by two units of the spatial dimension in the
electron-hole pairing in the lowest Landau level (LLL) with energy $E=0$\,
\cite{catal,Khvesh,Gorb}. As we discuss below, there is however an essential
difference between the QHE's in these two systems. While the pairing forces in
monolayer graphene lead to a relativistic-like scaling $\Delta E \sim \sqrt{|eB|}$
for the dynamical gap, in bilayer graphene, such a scaling
should take place only
for strong magnetic fields, $B \gtrsim B_{thr}$, where the threshold magnetic
field is estimated as $B_{thr} \sim 30{\mbox T}$ (see Sec. \ref{3b}).
For $B \lesssim B_{thr}$, a nonrelativistic-like scaling $\Delta E\sim |eB|$
is realized in the bilayer. The origin of this phenomenon is very different forms
of the polarization function in monolayer graphene and bilayer one that in turn
is determined by the different dispersion relations for quasiparticles in these
two systems.

The polarization function is one of the major players in the
QHE in bilayer graphene. As will be shown below, its role is important because
it is proportional to the large mass of quasiparticles,
{$m \sim 10^{-2}m_e \sim 10^8{\mbox K}/c^{2}\gg \hbar^{2}/e^2l$
unless $B \gtrsim 30$T,
which leads to strong screening.} 

Using the random phase approximation in the analysis of the gap
equation, we found two competing solutions: I) a ferromagnetic (spin splitting) solution,
and II) a layer asymmetric solution, actively discussed in the literature. Studying how
the energy gaps of these solutions depend 
{on the longitudinal component $B_{\parallel}$
of the magnetic
field at a fixed value of the transverse component $B_{\perp}$,}
we found that {while the gap of the solution I increases with
$B_{\parallel}$, the gap of the solution II decreases as
$B_{\parallel}$ increases.} Comparing this behavior
with that observed in experiment in Ref. \onlinecite{FMY} and calculating the energy
density of the ground states for these solutions, we come to the following scenario.
While at low magnetic fields, the layer asymmetric solution II is realized
with the energy gap $\Delta E\sim 10 B[{\mbox T}]{\mbox K}$, there
exists a first order phase transition to the ferromagnetic phase corresponding
to the solution I at some critical value $B_{cr}$.
The experiment \cite{FMY} implies that the value of $B_{cr}$ satisfies
$B_{cr} \gtrsim 10$T for $B_{\parallel}= 0$.
The phase diagram in the plane $(\tilde{\Delta}_0,B)$, where $\tilde{\Delta}_0$ is
a top-bottom gates voltage imbalance, is described.

The paper is organized as follows. In Sec. \ref{2}, the Hamiltonian of the model,
its symmetries, and order parameters are described. In Sec. \ref{3}, by using the
Baym-Kadanoff formalism \cite{BK}, the gap
equation for the quasiparticle propagator including the polarization function
is derived and the properties of the polarization function are described.
In Sec. \ref{4a}, the properties of the solutions of the gap equations
and the phase diagram of the model are discussed. In Sec. \ref{4b}, we
compare our results with experiment. In Sec. \ref{5}, we
summarize the main results of the paper.
In appendix \ref{A}, a detailed derivation of the polarization function
in a magnetic field in bilayer graphene is presented.

\section{Model}
\label{2}

\subsection{Hamiltonian}
\label{2a}

The free part of the effective low energy Hamiltonian of
bilayer graphene is \cite{McC}:
\be
H_0 = - \frac{1}{2m}\int
d^2x\Psi_{Vs}^+(x)\left( \begin{array}{cc} 0 & (\pi^{\dagger})^2\\ \pi^2 & 0
\end{array} \right)\Psi_{Vs}(x), \label{free-Hamiltonian}
\ee
where $\pi=\hat{p}_{x_{1}}+i\hat{p}_{x_{2}}$ and the canonical momentum
$\hat{\mathbf{p}} = -i\hbar\bm{\nabla}+ {e\mathbf{A}}/c$ includes the
vector potential $\mathbf{A}$ corresponding to the external magnetic field
$\mathbf{B}$. Without magnetic field, this Hamiltonian generates the spectrum
$E=\pm \frac{p^2}{2m}$, $m= \gamma_1/2v_{F}^2$, where the Fermi velocity $v_F
\simeq c/300$ and $\gamma_1 \approx 0.34-0.40$eV. The two component spinor
field $\Psi_{Vs}$ carries the valley $(V= K, K^{\prime})$ and spin
$(s = +, -)$ indices. We will use the standard convention:
$\Psi_{Ks}^T=(\psi_A{_1}, \psi_B{_2})_{Ks}$
whereas $\Psi_{K^{\prime}s}^T = (\psi_B{_2},
\psi_A{_1})_{K^{\prime}s}$. Here $A_1$ and $B_2$ correspond to
those sublattices in the layers 1 and 2, respectively, which, according to
Bernal $(A_2-B_1)$ stacking,
are relevant for the low energy dynamics. The effective Hamiltonian
(\ref{free-Hamiltonian}) is valid for magnetic fields $1T < B < B_{thr}$. For
$B < 1T$, the trigonal warping should be taken into account \cite{McC}. For $B
> B_{thr}$, a monolayer like Hamiltonian with linear dispersion should be used.

The Zeeman and Coulomb interactions plus
a top-bottom gates voltage imbalance $\tilde{\Delta}_0$
in bilayer graphene are described as
(henceforth we will omit indices $V$ and $s$ in the field $\Psi_{Vs}$):
\ba
H_{int}&=&
\mu_{B}B\int\hspace{-1.0mm} d^2x \Psi^+(x)\sigma^3\Psi(x) +
\frac{e^2}{2\kappa}\int\hspace{-1.0mm}
d^3xd^3x^{\prime}\frac{n(\mathbf{x})n(\mathbf{x}^{\prime})}{|\mathbf{x}-
\mathbf{x}^{\prime}|} + \tilde{\Delta}_0 \int d^2x \Psi^+(x)\xi\tau_3\Psi(x)\,,
\label{interaction}
\ea
where $\mu_B$ is the Bohr magneton, $\sigma^3$ is a spin matrix,
$\kappa$ is the dielectric constant, and
$n(\mathbf{x})=\delta(z-\frac{d}{2})\rho_1(x)+\delta(z+\frac{d}{2})\rho_2(x)$
is the three dimensional charge density ($d \simeq 0.35$nm is the distance
between the two layers). The Pauli matrix $\tau^3$ in the
voltage imbalance term acts on layer components,
and $\xi = \pm 1$ for the valleys $K$ and $K^{\prime}$, respectively.

{Integrating over $z$ and $z^{\prime}$ in this equation, one can rewrite
$H_{int}$ as
\ba
H_{int}&=& \mu_{B}B\hspace{-1.0mm} \int\hspace{-1.0mm} d^2x \Psi^+(x)
\sigma^3\Psi(x)\nonumber +
\frac{1}{2}\int \hspace{-1.0mm}d^2xd^2x^{\prime}\left[
V(x-x^{\prime})\left(\rho_1(x)\rho_1(x^{\prime})+
\rho_2(x)\rho_2(x^{\prime})\right)\hspace{-1.0mm} +
2V_{12}(x-x^{\prime})\rho_1(x)\rho_2(x^{\prime})\right]\\
&+&
\tilde{\Delta}_0 \int d^2x \Psi^+(x)\xi\tau_3\Psi(x)\,.
\label{interaction1}
\ea
Here the potential $V(x)$ describes the intralayer interactions and, therefore,
coincides with the bare potential in monolayer graphene whose Fourier transform
is given by $\tilde{V}(k)={2\pi e^2}/{\kappa k}$. The potential $V_{12}$ describes
the interlayer electron interactions. Its Fourier transform is
$\tilde{V}_{12}(k)=({2\pi e^2}/{\kappa})({e^{-kd}}/{k})$.
The two-dimensional charge densities $\rho_1(x)$ and $\rho_2(x)$ are:
\be
\rho_1(x)=\Psi^+(x)P_1\Psi(x)\,,\quad \rho_2(x)=\Psi^+(x)P_2\Psi(x)\,,
\label{density}
\ee
where $P_1=\frac{1+\xi\tau^3}{2}$ and $P_2=\frac{1-\xi\tau^3}{2}$ are projectors
on states in the layers 1 and 2, respectively. When the polarization
effects are taken into account, the potentials $V(x)$ and $V_{12}(x)$ are replaced
by effective interactions $V_{eff}(x)$ and $V_{12\,eff}(x)$, respectively,
whose Fourier transforms are given in Eqs.(\ref{interaction-D}) and
(\ref{interaction-ND}) in appendix \ref{A}.

\subsection{Symmetries and order parameters}
\label{2b}

The Hamiltonian $H = H_0 + H_{int}$, with $H_0$ and $H_{int}$ in Eqs.
(\ref{free-Hamiltonian}) and (\ref{interaction1}),
describes the dynamics
at the neutral point (with no doping). Because of the projectors $P_1$ and
$P_2$ in charge densities (\ref{density}), the symmetry of the Hamiltonian $H$
is essentially lower than the symmetry in monolayer graphene.
{If both the Zeeman
and $\tilde{\Delta}_0$ terms are ignored, it is
$U^{(K)}(2)_S \times U^{(K^{\prime})}(2)_S\times
Z_{2V}^{(+)}\times Z_{2V}^{(-)}$, where $U^{(V)}(2)_S$ defines the $U(2)$ spin
transformations in a fixed valley $V = K, K^{\prime}$, and $Z_{2V}^{(s)}$
describes the valley transformation $\xi \to -\xi$ for a fixed spin $s = \pm$
(recall that in monolayer graphene the symmetry would be $U(4)$ \cite{Gorb}).}
The Zeeman interaction lowers this symmetry down to $G_2 \equiv U^{(K)}(1)_{+}
\times U^{(K)}(1)_{-} \times U^{(K^{\prime})}(1)_{+} \times U^{(K^{\prime})}
(1)_{-} \times Z_{2V}^{(+)}\times Z_{2V}^{(-)}$, where $U^{(V)}(1)_{s}$ is the
$U(1)$ transformation for fixed values of both valley and spin. Recall that the
corresponding symmetry in monolayer graphene is $G_1 \equiv U^{(+)}(2)_V \times
U^{(-)}(2)_V$, where $U^{(s)}(2)_V$ is the $U(2)$ valley transformations for a
fixed spin. Including the $\tilde{\Delta}_0$ term lowers the $G_2$ symmetry
further down to the $\bar{G}_2 \equiv
U^{(K)}(1)_{+}\times U^{(K)}(1)_{-} \times U^{(K^{\prime}) }(1)_{+} \times
U^{(K^{\prime})}(1)_{-}$.

Although the $G_1$ and $G_2$ symmetries are quite
different, it is noticeable that their spontaneous breakdowns
can be described by the same
QHF and MC order parameters. The point is that these $G_1$ and $G_2$ define the
same four conserved commuting currents whose charge densities (and four
corresponding chemical potentials) span the QHF order parameters (we use the
notations of Ref. \onlinecite{Gorb1}):
{\begin{eqnarray}
\label{mu}
\mu_s:\quad
\langle{\Psi^\dagger_s\Psi_s}\rangle &=&
\langle{\psi_{K  A_1 s}^\dagger\psi_{K A_1 s}
+\psi_{K^{\prime} A_1 s}^\dagger\psi_{K^{\prime}A_1 s}
 +\psi_{K B_2 s}^\dagger\psi_{K B_2 s}
+ \psi_{K^{\prime}B_2 s}^\dagger\psi_{K^{\prime} B_2 s}}\rangle\,,\\
\label{mu-tilde}
\tilde{\mu}_{s}:\quad
\langle{\Psi^\dagger_s\xi\Psi_s}\rangle &=&
\langle{\psi_{K  A_1 s}^\dagger\psi_{K A_1 s}
- \psi_{K^{\prime} A_1 s}^\dagger\psi_{K^{\prime}A_1 s}
 + \psi_{K B_2 s}^\dagger\psi_{K
B_2 s} - \psi_{K^{\prime}B_2 s}^\dagger\psi_{K^{\prime} B_2 s}}\rangle\,.
\end{eqnarray}
The order parameter (\ref{mu}) is the charge density for a fixed spin
whereas the order parameter (\ref{mu-tilde}) determines the charge-density
imbalance between the two valleys. The corresponding chemical potentials
are $\mu_s$ and $\tilde{\mu}_s$, respectively.
While the former order parameter
preserves the $G_2$ symmetry, the latter completely
breaks its discrete subgroup $Z_{2V}^{(s)}$.
Their MC cousins are
\begin{eqnarray}
\label{delta}
\Delta_s: \quad
\langle{\Psi^\dagger_s\tau_3\Psi_s}\rangle &=&
\langle{\psi_{K  A_1 s}^\dagger\psi_{K A_1 s}
-\psi_{K^{\prime} A_1 s}^\dagger\psi_{K^{\prime}A_1 s}
 -\psi_{K B_2 s}^\dagger\psi_{K
B_2 s}+ \psi_{K^{\prime}B_2 s}^\dagger\psi_{K^{\prime} B_2 s}}\rangle\,,\\
\label{delta-tilde}
\tilde{\Delta}_{s}:\quad
\langle{\Psi^\dagger_s\xi\tau_3\Psi_s}\rangle &=&
\langle{\psi_{K  A_1 s}^\dagger\psi_{K A_1 s}
+\psi_{K^{\prime} A_1 s}^\dagger\psi_{K^{\prime}A_1 s}
 -\psi_{K B_2 s}^\dagger\psi_{K
B_2 s} - \psi_{K^{\prime}B_2 s}^\dagger\psi_{K^{\prime} B_2 s}}\rangle\,.
\end{eqnarray}

{These order parameters can be rewritten in the form of Dirac mass
terms \cite{Gorb1}. The corresponding masses are $\Delta_s$ and
$\tilde{\Delta}_{s}$, respectively. While the order parameter (\ref{delta})
preserves the $G_2$, it is odd under time reversal \cite{Hald}. On the other hand, 
the order parameter (\ref{delta-tilde}),
connected with the conventional Dirac mass $\tilde{\Delta}$,
determines the charge-density
imbalance between the two layers \cite{McC}.} Like the QHF order parameter
(\ref{mu-tilde}), this mass term completely breaks the $Z_{2V}^{(s)}$
symmetry and is even under $\cal{T}$.} Let us emphasize that unlike a
spontaneous breakdown of continuous symmetries, a spontaneous breakdown of
the discrete valley symmetry $Z_{2V}^{(s)}$, with the order parameters
$\langle{\Psi^\dagger_s\xi\Psi_s}\rangle$
and $\langle{\Psi^\dagger_s\xi\tau_3\Psi_s}\rangle$,
is not forbidden by the Mermin-Wagner theorem at finite
temperatures in a planar system\cite{MW}.

Note that because of the
Zeeman interaction, the $SU^{(V)}(2)_S$ is explicitly broken, leading to a
spin gap. This gap could be dynamically strongly enhanced \cite{Aban}. In that
case, a quasispontaneous breakdown of the $SU^{(V)}(2)_S$ takes place.
The corresponding ferromagnetic phase is described by the chemical potential
$\mu_3 = (\mu_+ - \mu_-)/2$, corresponding to the
QHF order parameter $\langle\Psi^\dagger\sigma_3\Psi\rangle$, and by the mass
$\Delta_3 = (\Delta_+ - \Delta_-)/2$ corresponding to
the MC order parameter
$\langle\Psi^\dagger\tau_3\sigma_3\Psi\rangle$ \cite{Gorb1}.

\section{Gap equation}
\label{3}

\subsection{General remarks}
\label{3a}

In this section, in the framework of the Baym-Kadanoff formalism
\cite{BK}, and using the polarization function calculated in the random phase
approximation (RPA), we derive and analyze the gap equation for the LLL quasiparticle
propagator with the order parameters introduced above. Recall that in bilayer
graphene, the LLL includes both the $n=0$ and $n=1$ Landau levels (LLs), if the Coulomb
interaction is ignored \cite{McC}. Therefore there are sixteen parameters
$\mu_{s}(n)$, $\Delta_{s}(n)$, $\tilde{\mu}_{s}(n)$, and
$\tilde{\Delta}_{s}(n)$ with $n=0, 1$.

As will be shown below, including the polarization function in the description of
the LLL dynamics is necessary. The point is that this function is proportional
to a large mass of quasiparticles, 
{$m \sim 10^{-2}m_e \sim 10^8{\mbox K}/c^{2}\gg \hbar^{2}/e^2l$}
unless $B \gtrsim 30$T},
which leads to strong screening effects. 

It will be shown below that the region of
relevant values of wave vector $\mathbf{k}$ in the gap equation for the LLL states
is $0 < y \equiv \mathbf{k}^2l^2/2 \lesssim 1$. While at
small $y \ll 1$ the dominant contribution {(around 80\%)}
in the polarization function
comes from the transitions between
the LLL and the first higher LL with n=2, the number of the LLs providing
relevant contributions in this function
grows with increasing $y$ (for details, see the analysis in
appendix \ref{A}).

Last but not least, a characteristic scale in the bilayer dynamics in a magnetic
field is the cyclotron energy $\hbar \omega_c \simeq 25.5 B[{\mbox T}]{\mbox K}$.
The applicability of the LLL approximation for a quasiparticle
propagator in the gap equation implies that the LLL energy gaps should be smaller
than $\hbar \omega_c$. As we will see, this condition is fulfilled in bilayer
graphene.

\subsection{The analysis of the gap equation}
\label{3b}

{The effective action in the Baym-Kadanoff formalism in two-loop approximation
is a functional of the full Green's function $G$ and
has the form
\ba
\Gamma(G)&=&{-i}\,\mbox{Tr}\left[\mbox{Ln} G^{-1} +S^{-1}G-1\right] - \int
d^{3}ud^3u^{\prime}\left\{\frac{1}{2}\mbox{tr}
\left[G(u,u^{\prime})G(u^{\prime},u)\right]V_{eff}(u-u^{\prime}) +
\mbox{tr}\,[P_1\,G(u,u^{\prime})\,P_2\,G(u^{\prime},u)\,]
\right.\nonumber\\
&\times&\left.V_{IL}(u-u^{\prime})-\frac{1}{2}\mbox{tr}\left[G(u,u)\right]\mbox{tr}\left[G(u^{\prime},
u^{\prime})\right]V_{eff}(u-u^{\prime})-\mbox{tr}\,[\,P_1\,G(u,u)]\,
\mbox{tr}\,[\,P_2\,G(u^{\prime},u^{\prime})\,]\,V_{IL}(u-u^{\prime})
\right\}\, , \label{CJT-action-all}
\ea
where $u \equiv (t,\mathbf{r})$, $t$ is the time coordinate and $\mathbf{r} = (x,y)$,
$V_{IL}(u)=V_{12\,eff}(u)-V_{eff}(u)$ is the interlayer interaction, and the Fourier transforms
of $V_{eff}(u)$ and $V_{12\,eff}(u)$ are given in Eqs.(\ref{interaction-D}) and
(\ref{interaction-ND}) in appendix \ref{A}.
Note that while here the trace
\rm{Tr}, the logarithm, and the product $S^{-1}G$ are taken in the functional sense, the trace
\rm{tr} runs over spinor} and spin indices.

{The stationary condition $\delta\Gamma(G)/\delta G =0$
leads to the gap (Schwinger-Dyson) equation in mean field approximation,
which will be written
in the form convenient in the presence of a magnetic field:
\ba
G(u_1,u_2) &=& S(u_1,u_2) + i \int d^3u^{\prime}_1d^3u^{\prime}_2 \, S(u_1,u^{\prime}_1)\,
G(u^{\prime}_1,u^{\prime}_2)\,G(u^{\prime}_2,u_2)\,\,V_{eff}(u^{\prime}_1 - u^{\prime}_2)\nonumber\\
&+&i\,\int d^3u^{\prime}_1d^3u^{\prime}_2 \,
S(u_1,u^{\prime}_1)\,\left[\,P_1\,G(u^{\prime}_1,u^{\prime}_2)\,P_2+
P_2\,G(u^{\prime}_1,u^{\prime}_2)\,P_1\,\right]\,G(u^{\prime}_2,u_2)\,
V_{IL}(u^{\prime}_1 - u^{\prime}_2)\nonumber\\
& -&i\int d^3u^{\prime}_2\,S(u_1,u^{\prime}_2)\,\left\{\,\mbox{tr}\,[\,G(u_1,u_1)\,]\,\tilde{V}_{eff}(0)+
\left(P_1\,\mbox{tr}\,[P_2\,G(u_1,u_1)\,]\right.\right.\nonumber\\
 &+&\left.\left. P_2\,\mbox{tr}\,[\,P_1\,G(u_1,u_1)\,]\,
\right)\,\tilde{V}_{IL}(0)\,\right\}G(u^{\prime}_2,u_2),
\label{SDeq-magnetic-all}
\ea
where $\tilde{V}_{eff}(0)$ and $\tilde{V}_{IL}(0)$ are the Fourier transforms of
$V_{eff}(u)$ and $V_{IL}(u)$ taken at $\omega=\mathbf{k}=0$.}

{We will use the Landau gauge for a two dimensional vector potential,
$\mathbf{A_{\parallel}}=(0,B_{\perp}x)$,
where $B_{\perp}$ is the component of the
magnetic field $\mathbf{B}$ orthogonal to the $xy$ plane of graphene. Then,
the free Green's function $S(u_1,u_2)$ can be written as a product of a translation
invariant part $\tilde{S}(u_1 -u_2)$ times the
Schwinger phase factor \cite{catal,schwinger},
\be
S(u_1,u_2)=\exp\left(-i\frac{(x_1 + x_2)(y_1 -y_2)}{2l^{2}}\right)\tilde{S}(u_1 - u_2).
\ee
After extracting the Schwinger phase factor in the full propagator,
\be
G(u_1,u_2)=\exp\left(-i\frac{(x_1 + x_2)(y_1 -y_2)}{2l^{2}}\right)\tilde{G}(u_1 - u_2),
\label{coordinate-space-propagator}
\ee
and making the Fourier transform with respect to $t$, we get the following equation
for the translation invariant part $\tilde{G}$:
\ba
\tilde{G}(\Omega,\mathbf{r})&=&\tilde{S}(\Omega,\mathbf{r})
+i\int\frac{d\omega}{2\pi}\int d^2\mathbf{r^{\prime}_1} d^2\mathbf{r^{\prime}_2}\,
e^{i[(x - x^{\prime}_{2})y^{\prime}_{1}-
(y - y^{\prime}_{2})x^{\prime}_{1}]/2l^{2}}\tilde{S}(\Omega,\mathbf{r}- \mathbf{r^{\prime}_1})\nonumber\\
&\times&\left[
\tilde{G}(\omega,\mathbf{r^{\prime}_1}-\mathbf{r^{\prime}_2})\,
V_{eff}(\Omega-\omega,\mathbf{r^{\prime}_1}-\mathbf{r^{\prime}_2})
+\left(P_{1}\tilde{G}(\omega,\mathbf{r^{\prime}_1}-\mathbf{r^{\prime}_2})P_{2}+P_{2}\tilde{G}(\omega,
\mathbf{r^{\prime}_1}-\mathbf{r^{\prime}_2})P_{1}\right)V_{IL}
(\Omega-\omega,\mathbf{r^{\prime}_1}-\mathbf{r^{\prime}_2})\right]
\nonumber\\
&\times&\tilde{G}(\Omega,\mathbf{r^{\prime}_2})-i\int d^{2}\mathbf{r^{\prime}_2}\,
e^{i[xy^{\prime}_{2} - yx^{\prime}_{2}]/2l^{2}}
\tilde{S}(\Omega,\mathbf{r} - \mathbf{r^{\prime}_2})
\left\{\,\mbox{tr}\,[\tilde{G}(0)\,]\,\tilde{V}_{eff}(0)\right.\nonumber\\
&+&\left.
\left(P_1\,\mbox{tr}\,[P_2\,\tilde{G}(0)\,]+P_2\,\mbox{tr}\,[\,P_1\tilde{G}(0)\,]\,
\right)\tilde{V}_{IL}(0)\,\right\}\tilde{G}(\Omega,\mathbf{r^{\prime}_2}),
\label{eq:fullpropagator}
\ea
where $\mathbf{r} = \mathbf{r}_1 - \mathbf{r}_2$.}

The translation invariant part of the free propagator
can be expanded over the LLs (compare with Refs. \onlinecite{Gorb1,catal}):
\ba
\tilde{S}_{\xi s}(\mathbf{r};\omega)&=&\frac{1}{2\pi l^{2}}
\exp\left(-\frac{\mathbf{r}^{2}}{4l^{2}}\right)\sum\limits_{n=0}^{\infty}
\frac{1}{(\omega+ i\delta{\rm sgn\omega} +\bar{\mu}_s)^{2}-E^{2}_{n}}\left\{(\omega+\bar{\mu}_s
+\xi\tau_{3}\tilde{\Delta}_{0})
\left[P_{-}L_{n}\left(\frac{\mathbf{r}^{2}}{2l^{2}}\right)+P_{+}L_{n-2}
\left(\frac{\mathbf{r}^{2}}{2l^{2}}\right)\right]\right.\nonumber\\
&+&\left.\frac{\hbar^{2}}{2m l^{4}}L^{2}_{n-2}\left(\frac{\mathbf{r}^{2}}{2l^{2}}\right)
\left(\begin{array}{cc}0& (x-i y)^{2} \\(x+i y)^{2}& 0\end{array}\right)\right\},
\label{free-propagator}
\ea
where $P_{\pm}=(1\pm\tau_{3})/2$, $E_{n}=
\sqrt{\hbar^2\omega_{c}^{2}n(n-1)+\tilde{\Delta}_{0}^{2}}$,
$\omega_{c}= eB/m c$ is the cyclotron frequency,
and the bare electron chemical potential $\bar{\mu}_{s}=\mu_{0}-s Z$
includes the Zeeman energy $Z\simeq\mu_{B}B=0.67\,B[{\mbox T}]{\mbox K}$
(the conventional chemical potential $\mu_0$, responsible for doping,
is included for generality). The functions
$L^{\alpha}_{n}(x)$ are generalized Laguerre polynomials,
and by definition, $L_n(x)= L^{0}_n(x)$, $L^{\alpha}_{-2}(x)=
L^{\alpha}_{-1}(x)\equiv0$.

For the LLL
with $n = 0, 1$, expression (\ref{free-propagator}) takes a simple form:
\be
\tilde{S}_{\xi s}(\mathbf{r};\omega)=\frac{1}{2\pi l^{2}}\exp\left(-\frac{\mathbf{r}^{2}}
{4l^{2}}\right)\left[L_{0}\left(\frac{\mathbf{r}^{2}}{2l^{2}}\right)+
L_{1}\left(\frac{\mathbf{r}^{2}}{2l^{2}}\right)\right]{S}_{\xi s}(\omega)P_{-},
\label{free-propagator-LLL}
\ee
where
\be
{S}_{\xi s}(\omega)=\frac{1}{\omega+\bar{\mu}_{s}+\xi\tilde{\Delta}_{0} + i\delta{\rm sgn\omega}}.
\ee

{Motivated by expression (\ref{free-propagator-LLL}) for the free propagator in the
LLL approximation, we will use the following ansatz for the full propagator with the
parameters $\mu_s(n)$, $\tilde{\mu}_{s}(n)$, $\Delta_s(n)$,
and $\tilde{\Delta}_s(n)$ related to the order parameters in Eqs. (\ref{mu}) -- (\ref{delta-tilde}):
\be
\tilde{G}_{\xi s}(\mathbf{r};\omega)=\frac{1}{2\pi l^{2}}\exp\left(-\frac{\mathbf{r}^{2}}
{4l^{2}}\right)\left[G_{\xi s0}(\omega)L_{0}\left(\frac{\mathbf{r}^{2}}{2l^{2}}\right)+
G_{\xi s1}(\omega)L_{1}\left(\frac{\mathbf{r}^{2}}{2l^{2}}\right)\right]P_{-},
\label{ansatz}
\ee
where
\be
G_{\xi s n}(\omega)=\frac{1}{\omega-E_{\xi ns}+ i\delta{\rm sgn\omega}},
\label{G-expression}
\ee
and
\be E_{\xi ns}= -(\mu_{s}(n)+ \Delta_{s}(n))+ \xi(\tilde{\mu}_{s}(n) -
\tilde{\Delta}_{s}(n)),\quad n=0,1,
\label{disp}
\ee
are the energies of the LLL states depending on the order parameters $\mu_{s}(n),
\tilde{\mu}_{s}(n),\Delta_{s}(n),\tilde{\Delta}_{s}(n)$.

{Inserting the ansatz (\ref{ansatz}) into Eq.(\ref{eq:fullpropagator}) and using the
orthogonality of the Laguerre polynomials,
we derive the following system of equations  for
the functions $G_{\xi sn}(\omega)$:
\ba
G^{-1}_{\xi s0}(\Omega)&=&S^{-1}_{\xi s}(\Omega)-i\int\frac{d\omega\,d^{2}k}{(2\pi)^{3}}\,
[G_{\xi s0}(\omega)+ G_{\xi s1}(\omega)\mathbf{k}^{2}l^{2}/2]\,
 e^{-\mathbf{k}^{2}l^{2}/2}\tilde{V}_{eff}\left(\Omega-\omega,|\mathbf{k}|\right)\nonumber\\
&+&\frac{1}{4\pi l^2}\,\left(\frac{1+\xi}{2}A_1 +\frac{1-\xi}{2}A_2\,
\right)\tilde{V}_{IL}(0),
\label{gap-equation-g-2}\\
G^{-1}_{\xi s1}(\Omega)&=&S^{-1}_{\xi s}(\Omega)-i\int\frac{d\omega\,d^{2}k}{(2\pi)^{3}}\,
[G_{\xi s0}(\omega)\mathbf{k}^{2}l^{2}/2+G_{\xi s1}(\omega)
(1-\mathbf{k}^{2}l^{2}/2)^{2}]\,e^{-\mathbf{k}^{2}l^{2}/2}\tilde{V}_{eff}\left(
\Omega-\omega,|\mathbf{k}|\right)\nonumber\\
&+&\frac{1}{4\pi l^2} \left(\frac{1+\xi}{2}A_1 +\frac{1-\xi}{2}A_2
\right)\tilde{V}_{IL}(0). \label{gap-equation-g-3}
\ea
Here  $A_1 =
\sum_{n,s}\,\mbox{sgn}(\,E_{- ns})$\,, $A_2 =\sum_{n,s}\,\mbox{sgn}(\,E_{+ ns})$.
The second and third terms on
right hand sides of Eqs.(\ref{gap-equation-g-2}) and (\ref{gap-equation-g-3})
describe the Fock and Hartree interactions, respectively. Note that because
for the LLL states only the component $\psi_{B_2s}$ $(\psi_{A_1s})$ of the wave
function at the $K (K^{\prime})$ valley is nonzero, their energies depend only
on the eight independent combinations of the QHF and MC parameters shown in
Eq. (\ref{disp}).}

{As is shown in appendix \ref{A}, neglecting the dependence on $d$ in the function
$\tilde{V}_{eff}(\omega,k)$ describing the exchange interactions,
one gets
\be
\tilde{V}_{eff}(\omega,k)=\frac{2\pi e^2}{\kappa}\,\frac{1}{ k
+\frac{4\pi e^2}{\kappa}\Pi (\omega,{\bf k}^2)}
\label{V-D}
\ee
with $\Pi(\omega,k^{2})\equiv \Pi_{11}(\omega,\mathbf{k})+\Pi_{12}(\omega,\mathbf{k})$,
where the polarization function $\Pi_{ij}$ describes electron densities
correlations on the layers $i$ and $j$ in a magnetic field (see Eqs. (\ref{coulomb-interaction}),
(\ref{p11}), and (\ref{p12})). As to the Hartree interactions, it is
(see Eq.(\ref{interlayer-interaction})):
\be
\tilde{V}_{IL}(\omega=0,{k}=0)=-\frac{2\pi e^{2}d}{\kappa_{eff}},\quad
\kappa_{eff}=\kappa+2\pi e^{2}d(\Pi_{11}(0)-\Pi_{12}(0)).
\label{interlayer-interaction1}
\ee
It is estimated in
appendix \ref{A} that the value of the dynamical part of $\kappa_{eff}$, i.e., $\kappa_{eff} - \kappa$,
varies in the interval $1\div 4$.}

We utilize the frequency independent order parameters $\mu,\tilde{\mu},\Delta,\tilde{\Delta}$
and take the external frequency $\Omega=0$ in Eqs. (\ref{gap-equation-g-2}) and
(\ref{gap-equation-g-3}). The static approximation for the polarization function will be
used, $\Pi(\omega,{\bf k}^2) \to \Pi(0,{\bf k}^2)$. {As a justification of the latter,
we present the following argument. Let us assume
that the main contribution in the integrals over $\omega$ in 
Eqs.(\ref{gap-equation-g-2}) and
(\ref{gap-equation-g-3}) comes from the {pole} terms in 
the functions $G_{\xi s n}(\omega)$ (see
Eq.(\ref{G-expression})). The contribution of each pole
in the polarization function  has the form
$\Pi (\omega=E_{\xi n s},{\bf k}^2)$, and the dependence on $E_{\xi n s}$ enters
through $(E_{\xi n s}/\hbar\omega_{c})^{2}$, where the
cyclotron energy $\hbar\omega_{c} \simeq 25.5 B[{\mbox T}]{\mbox K}$.
As will be shown below, the ratio $(E_{\xi n s}/\omega_{c})^{2}$ is small,
$\sim 0.15$, and, therefore,
can be neglected in Eq.(\ref{Pi:frequency}), which leads to a
static polarization function $\Pi(0,{\bf k}^2)$.

It is convenient to rewrite the static polarization $\Pi(0,{\bf k}^2)$ in the form
$\Pi = (m/{\hbar}^2)\tilde{\Pi}(y)$, where both $\tilde{\Pi}$ and
$y \equiv \mathbf{k}^2l^2/2$ are dimensionless.
The function $\tilde{\Pi}(y)$ was expressed in terms of the sum over
all the Landau levels (see Eq. (\ref{Pi:static}) in appendix \ref{A})
and was analyzed both analytically and numerically. At $y\ll 1$,
$\tilde{\Pi}(y) \simeq 0.55y$ and its derivative $\tilde{\Pi}'$ changes
from $0.55$ at $y=0$ to 0.12 at $y=1$. At large $y$ it
approaches a zero magnetic field value, $\tilde{\Pi}(y)\simeq \ln4/\pi$
(see Fig. \ref{fig1}) \cite{footnote}.
\begin{figure}[htp]
\includegraphics[width=6.0cm]{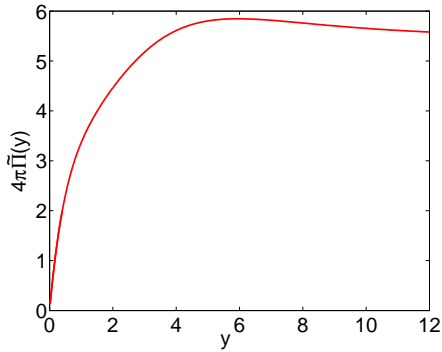}
\caption{ The static polarization function $4\pi\tilde{\Pi}(y)$. }
\label{fig1}
\end{figure}

Because of the Gaussian factors $e^{-\mathbf{k}^{2}l^{2}/2} = e^{-y}$ in Eqs.
(\ref{gap-equation-g-2}) and (\ref{gap-equation-g-3}), the relevant region in
the integrals in these equations is $0 < y \lesssim 1$.
The crucial point in the analysis
is that the region where the bare Coulomb term $k$ in the
denominator of ${V}_{eff}(k)\equiv {V}_{eff}(0,k)$ (\ref{V-D}) dominates is
very small, $0 < y \lesssim 10^{-3}B$[T].The main reason of that is a large
mass of quasiparticles, 
$m \sim 10^{-2}m_e \sim 10^8{\mbox K}/c^{2}\gg \hbar^{2}/e^2l$.
{The last inequality takes place unless $B \gtrsim 30$T.} 
As a result, the polarization function term
dominates in ${V}_{eff}(k)$ that leads to ${V}_{eff}(k)$ of the form
${V}_{eff}(k)= C(y)\hbar^2/ml^2k^2$.
The factor $\hbar^2/ml^2k^2$ has the same $k$ dependence as
the Coulomb potential in two dimensions, and the
factor $C(y)$ describes its smooth modulations at $0\leq y \lesssim 1$
(see Fig. \ref{fig1}).
It is unlike the case of monolayer graphene where the effective interaction is
proportional to $1/k$.

{By using the change of variables ${\bf k} \to l{\bf k}$} in
Eqs.(\ref{gap-equation-g-2}) and (\ref{gap-equation-g-3}), one can see
that $|eB|$ occurs as an overall factor in the front of the integrals in
these equations. {The latter leads} to the
scaling $\Delta E \sim |eB|$ for the
dynamical energy gap, and not $\Delta E \sim \sqrt{|eB|}$ taking place in
monolayer graphene \cite{Gorb1,QHF,MC} (see Sec. \ref{4a} below).

As shown in appendix \ref{A}, the contribution of the LLL with
$n = 0, 1$ in the polarization function is identically zero.
At $y \ll 1$, the main contribution (around $80\%$)
comes from the transitions between the LLL
and the first higher LL with $n = 2$. With increasing $y$, the number of higher LLs providing
relevant contributions in the polarization function grows.

As to the condition of the applicability of this low energy model,
according to Ref. \onlinecite{McC}, it is determined by the relation
$\hbar \omega_c\sqrt{n(n-1)} \leq\gamma_1/4$. Its left-hand side is nonzero
for $n \ge 2$ and increases linearly with $B$. Taking $n=2$ and the sign
of equality in this relation, we find the threshold magnetic field
$B_{thr}=\frac{45}{\sqrt{2}}\,T \approx 32\,T$ that
determines the upper limit for the values of $B$ for
which the low energy model is applicable.

{With the static polarization function, the
integration over the frequency $\omega$ in Eqs.(\ref{gap-equation-g-2}) and
(\ref{gap-equation-g-3}) can be performed explicitly, and we get a system of
algebraic equations for the energies $E_{\xi n s}$ in Eq. (\ref{disp}):
\ba
-E_{\xi 0s}&=&{\mu}_{0}-s Z+\xi\tilde{\Delta}_{0}-\frac{1}{2ml^{2}}\left[{\rm sgn}
\left(E_{\xi 0s}\right)I_{1}(x)+{\rm sgn}\left(E_{\xi 1s}\right)I_{2}(x)\right]\nonumber\\
&+&\frac{1}{4\pi l^{2}}\left[(A_{1}+A_{2})\,V_{eff}(0)+
\left(\frac{1-\xi}{2}A_{2}+\frac{1+\xi}{2}A_{1}\,
\right)\,{V}_{IL}(0)\right],
\label{system-eq1}\\
-E_{\xi 1s}&=&{\mu}_{0}- s Z+\xi\tilde{\Delta}_{0}-\frac{1}{2ml^{2}}\left[{\rm
sgn}
\left(E_{\xi 0s}\right)I_{2}(x)+{\rm sgn}\left(E_{\xi 1s}\right)I_{3}(x)\right]\nonumber\\
&+&\frac{1}{4\pi l^{2}}\left[(A_{1}+A_{2})\,{V}_{eff}(0)+
\left(\frac{1-\xi}{2}A_{2}+\frac{1+\xi}{2}A_{1}
\right)\,{V}_{IL}(0)\right],
\label{system-eq2}
\ea
where the quantities $I_{i}(x)$ are
\begin{equation}
I_i(x)=\int_0^{\infty} \frac{dy\,f_i(y)\,e^{-y}}{\kappa \sqrt{xy} + 4\pi
\tilde{\Pi}(y)}
\label{integralsI-i}
\end{equation}
with $f_i(y)=(1,\,y,\,(1-y)^2)$ for $i=1,2,3$, respectively.
Here the dimensionless
variable $x=2\hbar^{4}/e^{4}m^{2}l^{2}=(4\hbar\omega_{c}/\alpha^{2}\gamma_{1})
(v_{F}/c)^{2}\simeq 0.003 B[{\mbox T}]$, where $\alpha=1/137$ is the
fine-structure constant and
we used the values $\gamma_{1}=0.39
\mbox{eV}$, $\hbar\omega_{c}=\hbar^{2}/ml^{2}=2.19 B[{\mbox T}]\mbox{meV}$,
$v_{F}=8.0\times10^{5}$m/s
(see Ref. \onlinecite{McC}).

\section{Solutions and phase diagram: theory and experiment}
\label{4}

\subsection{Properties of solutions}
\label{4a}

In this section the solutions of Eqs. (\ref{system-eq1}) and (\ref{system-eq2})
and the phase diagram of the system these solutions lead to will be described.
If the Zeeman term is ignored, the equations for parameters with
different spin indices coincide. If the voltage imbalance
$\tilde{\Delta}_0$ term is absent, these equations are also invariant with respect
to the permutation of layer indices ($\xi\to -\xi$)
[note that $A_{1}\leftrightarrow A_{2}$ under the change $\xi\to -\xi$].
Clearly, these symmetries of the gap equations are due to the
$SU^{(K)}(2)_S \times SU^{(K^{\prime})}(2)_S\times
Z_{2V}^{(+)}\times Z_{2V}^{(-)}$ symmetry of the bilayer Hamiltonian
discussed in Sec. \ref{2b}
(note that if the interlayer Coulomb interaction
term ${V}_{IL}$ were absent, we would have the $U(4)$ symmetry group,
as in monolayer graphene).

Due to the Zeeman and $\tilde{\Delta}_0$ terms, these
equations are inhomogeneous. It is natural to expect that the lowest energy
solution will have the sign correlating with the sign of inhomogeneous terms
(solutions with different signs are degenerate in the case of homogeneous
equations). Without loss of generality, we can assume that $\tilde{\Delta}_0$ is
positive.

At the neutrality point ($\mu_0 = 0$ and  $A_1+A_2=0$), we found two
competing solutions of these equations: I) a ferromagnetic (spin splitting) solution, and II)
a layer asymmetric solution, actively discussed in the literature.
The energy (\ref{disp}) of the LLL states of the solution I equals:
\be
E^{(I)}_{\xi ns}=
s\left(Z+\frac{\hbar^{2}}{2ml^2}F_{n}(x)\right) - \xi\tilde{\Delta}_0 \,, \label{solution-I}
\ee
where $F_{0}(x)=I_{1}(x)+I_{2}(x)$ and $F_{1}(x)=I_{2}(x)+I_{3}(x)$
with $I_i$ in Eq. (\ref{integralsI-i}).
The solution exists for $\tilde{\Delta}_0<{Z}+\frac{\hbar^{2}}{2m l^{2}}F_1(x)$.
{Since $A_{1}=A_{2}=0$ in this solution, the Hartree interaction does not 
contribute in $E^{(I)}_{\xi ns}$.
Note that the
dynamical term $(\hbar^{2}/2ml^2)F_{n}(x)$ in Eq. (\ref{solution-I}) can be rewritten as
$(\hbar|eB|/2mc) F_{n}(x)$, where $F_{n}(x)$ depends on $B$ logarithmically for $x \ll 1$.}

{The energy (\ref{disp}) of the LLL states of the solution II is different:}
\ba
 E^{(II)}_{\xi ns}&=&
s Z - \xi\left(\tilde{\Delta}_0 +
\frac{\hbar^{2}}{2ml^2}F_{n}(x) - \frac{2e^2 d}{\kappa_{eff} l^2}\right)\,. \label{solution-II}
\ea
The last term in the parenthesis is the Hartree one, and the solution exists for
$\tilde{\Delta}_0>\frac{2e^{2}d}{\kappa_{eff} l^{2}}+Z-\frac{\hbar^{2}}{2m l^{2}}
F_1(x)$.} For illustrative purpose, in suspended bilayer graphene, with $\kappa \sim 1$,
we will use $\kappa_{eff} = 4$ (see Eq. (\ref{interlayer-interaction1})).

The energy density of the ground state for these solutions is ($a = I, II$):
\ba
\epsilon^{(a)}=-\frac{1}{8\pi l^2}
\sum_{\xi=\pm}\sum_{s=\pm}\sum_{n=0,1}\left[|E_{\xi ns}^{(a)}|
 + (-s\,0.67B+\xi\tilde{\Delta}_0)\,\mbox{sgn}\,E_{\xi ns}^{(a)}\right].
 \label{dens}
\ea
It is easy to check that for balanced bilayer ($\tilde{\Delta}_0 = 0$)
the solution I is favorite. There are two reasons of that: the presence of the Zeeman term and the
capacitor like Hartree contribution in the energy $E^{(II)}_{\xi ns}$
in the solution II.
\begin{figure}[ht]
\includegraphics[width=7.0cm]{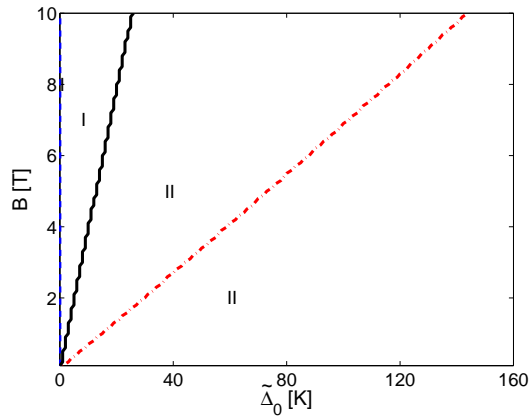}
\caption{ The phase diagram in the $(\tilde{\Delta}_0,B)$ plane at $B_\parallel =0$.
Here the effective $\kappa_{eff}=4.$}
\label{fig2}
\end{figure}
In Fig.~\ref{fig2}, the phase diagram on the plane $(\tilde{\Delta}_0,B)$,
at $B_\parallel =0$, is
presented. The I (II) area is that where the solution I (solution II)
is favorite. The two dashed lines compose the boundary of the region where the
two solutions coexist (the solution I does not exist to the right of the (red)
dashed line in the region II, while the solution II does not exist
to the left of the (blue) dashed line in the region I).
The black bold line is a line of a phase transition between the phases
I and II. Because the solutions coexist in the region around that line,
the phase transition is a first order one.
The equation for the critical value $B_{cr}$ has a simple form,
$B_{cr}[{\mbox T}] \simeq 0.4 \tilde{\Delta}_0$[K].

It is noticeable that for any fixed value of $B\, (\tilde{\Delta}_0)$,
there are sufficiently large values of $\tilde{\Delta}_0$\, (B),
at which the solution I (solution II) does not exist at all. It is because a
voltage imbalance (Zeeman term) tends to destroy the solution I (solution II).

\begin{figure}[ht]
\centering{
\includegraphics[width=7.0cm]{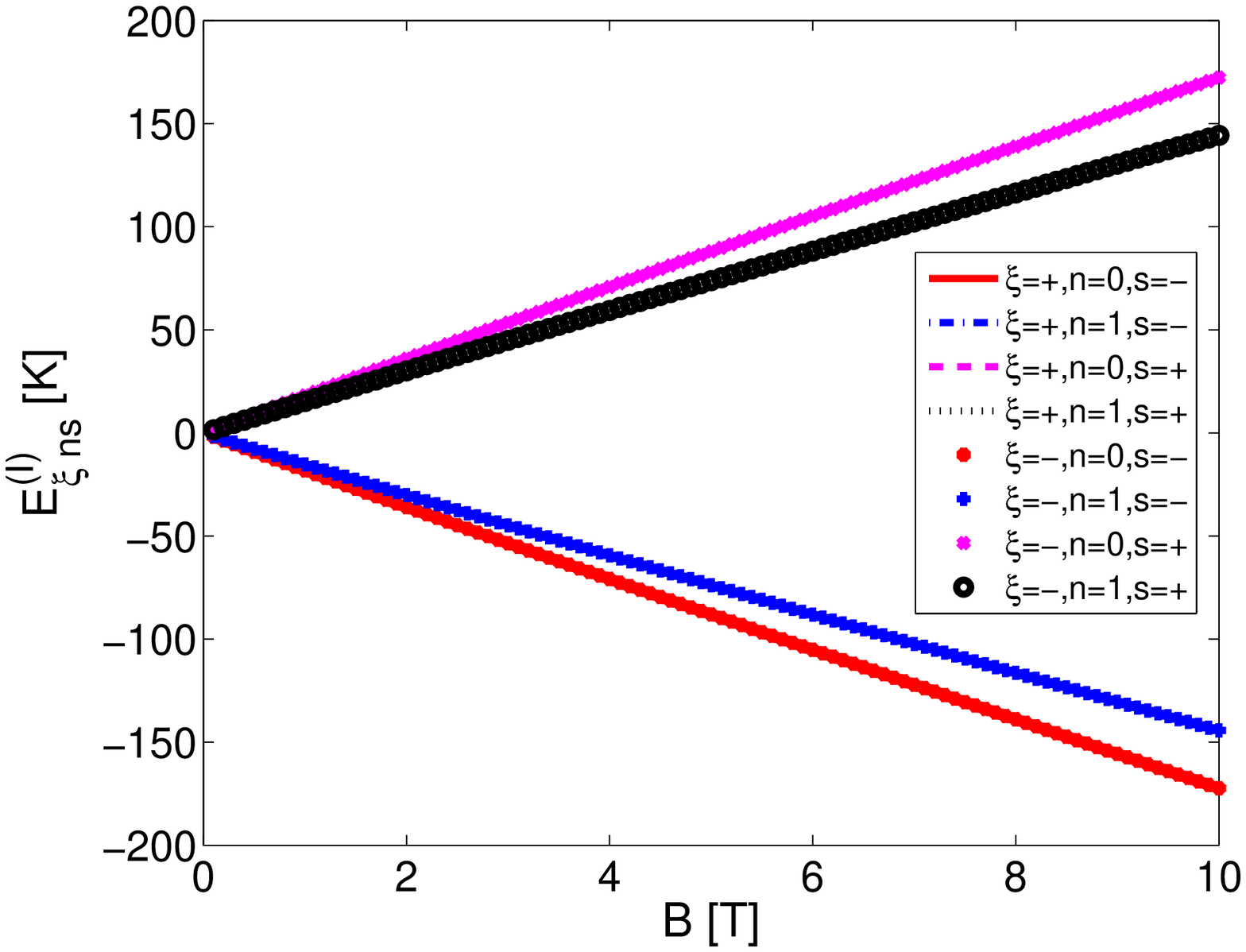}
\includegraphics[width=7.0cm]{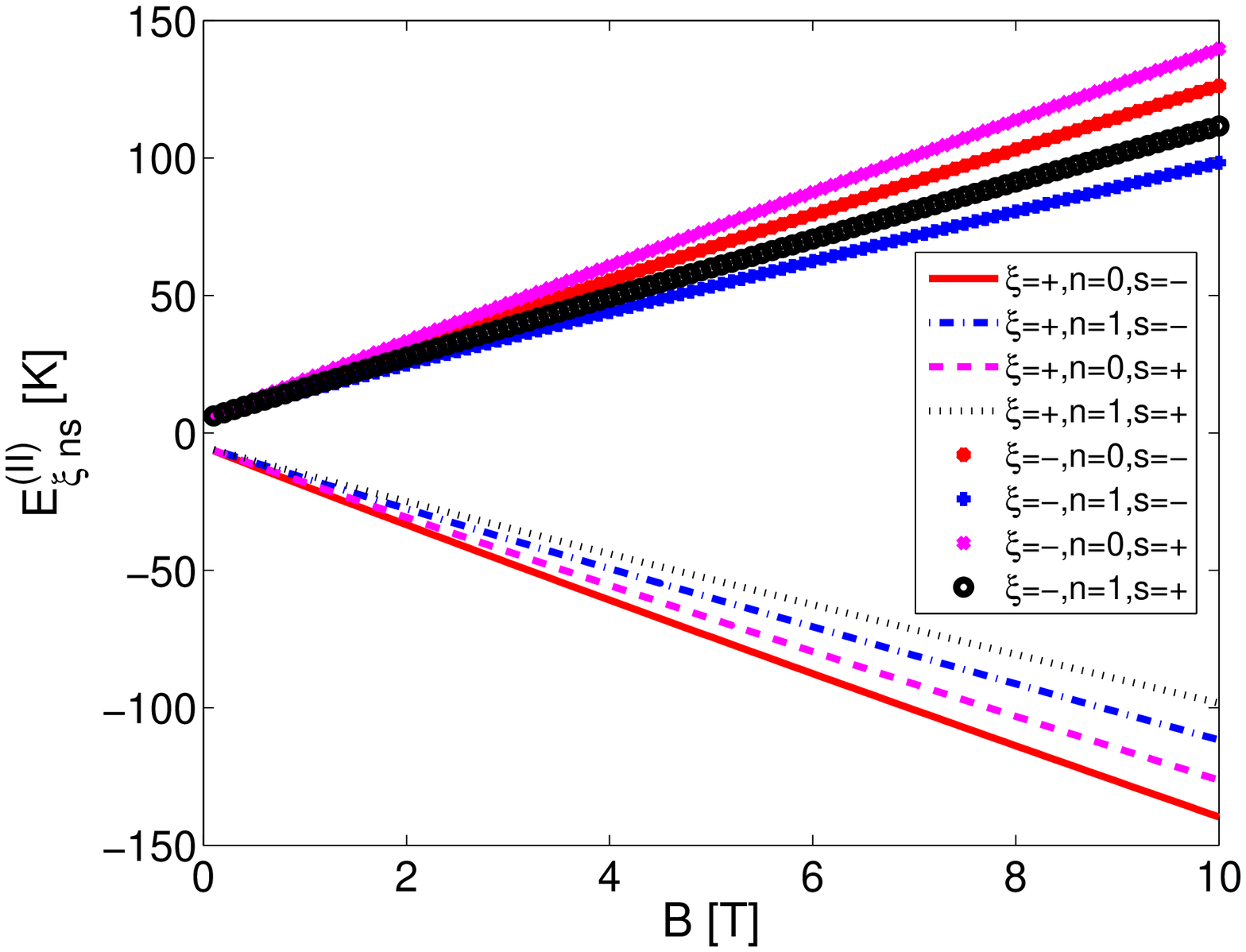}
}
\caption{The LLL energies of the solutions I (left panel) and II (right panel)
as functions of $B$ with $B_{\parallel}=0$.
Here $\tilde{\Delta}_{0}=0$ and $\tilde{\Delta}_{0}=5 \mbox{K}$ for solution I
and solution II, respectively.}
 \label{fig3}
\end{figure}

For $\tilde{\Delta}_0 = 0$, the dependence of the LLL energies
$E^{(I)}_{\xi ns}$ of the solution I on $B$, at $B_{\parallel}=0$,
is shown on the left panel
in Fig.~\ref{fig3} (the LLL states with opposite $\xi$ remain
degenerate in this solution). The perfectly linear form of this dependence is evident.
Also, the degeneracy between the states of the $n=0$ LL and those of the $n =
1$ LL is removed. The energy gap corresponding to the $\nu = 0$ plateau is
$\Delta E^{(I)} = (E_{\xi1 + }^{(I)} - E_{\xi1 - }^{(I)})/2 \simeq 14.4 B [{\mbox
T}]$K.

{On the right panel in Fig.~\ref{fig3}, the
dependence of the LLL energies
of the solution II on $B$, at $B_{\parallel}=0$,
is shown for $\tilde{\Delta}_0= 5 \mbox{K}$.
It is also perfectly linear.
Unlike the
solution I, the LLL degeneracy is now completely removed. As to
the energy gap corresponding to the $\nu = 0$ plateau, it is
$\Delta E^{(II)} = (E_{- 1 - }^{(II)} - E_{+ 1 + }^{(II)})/2  \simeq 5{\mbox K} + 9.3
B [{\mbox T}]$K.}

Fig. \ref{fig4} illustrates how the energy gaps of the two solutions depend on the
longitudinal component of the magnetic field $B_{\parallel} = B \cos\alpha$ at
a fixed value of the transverse component $B_{\perp} = B \sin\alpha$. As one can see, while
the gap of the solution I {increases with $B_{\parallel}$, the
gap of the solution II decreases as $B_{\parallel}$ increases.} These properties
of course reflect the opposite roles of the Zeeman term in the dynamics of the solutions
I and II.

{Thus the results of the analysis of this subsection imply a possibility of
the following two scenarios. When the
top-bottom gates voltage imbalance $\tilde{\Delta}_0 = 0$, the ferromagnetic phase I
is favorite for all values of the magnetic field. For nonzero $\tilde{\Delta}_0$, the phase II
is realized for the values
magnetic fields up to the critical value $B_{cr}$, where a first order
phase transition to the ferromagnetic phase I takes place. As we will discuss in the next subsection,
the experiment in Ref. \onlinecite{FMY} clearly prefers the second scenario.}

What can be the origin of $\tilde{\Delta}_0$? As was pointed out in Ref. \onlinecite{FMY},
it could be generated due to disorder-induced differences in carrier density between
the top and bottom layers.
A more interesting possibility is that a relatively small $\tilde{\Delta}_0$ is a
dynamical parameter corresponding to
spontaneous breakdown of the discrete valley symmetry
$Z_{2V}^{(+)}\times Z_{2V}^{(-)}$ in bilayer graphene with no magnetic field.

\begin{figure}[ht]
\centering{
\includegraphics[width=7.0cm]{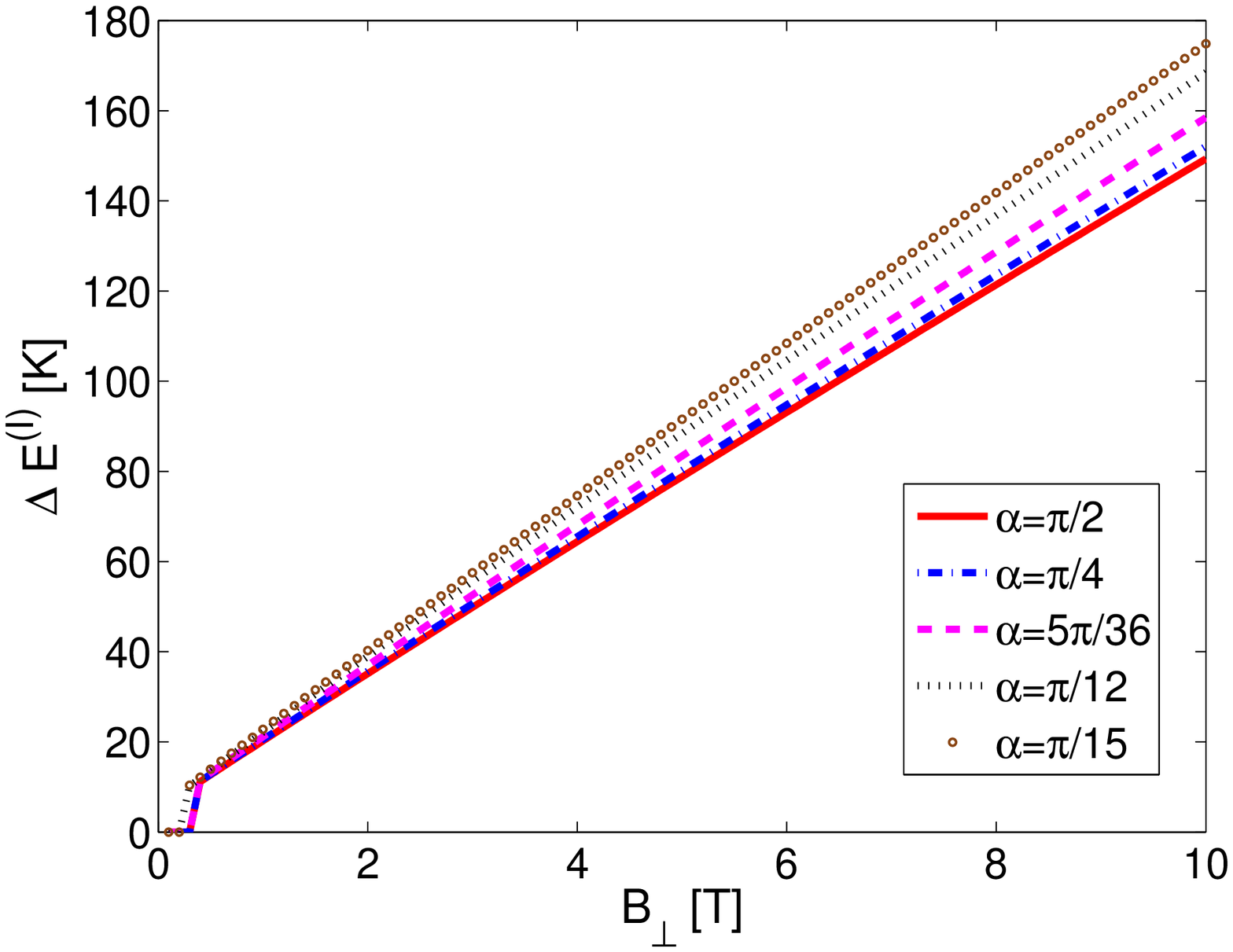}
\includegraphics[width=7.0cm]{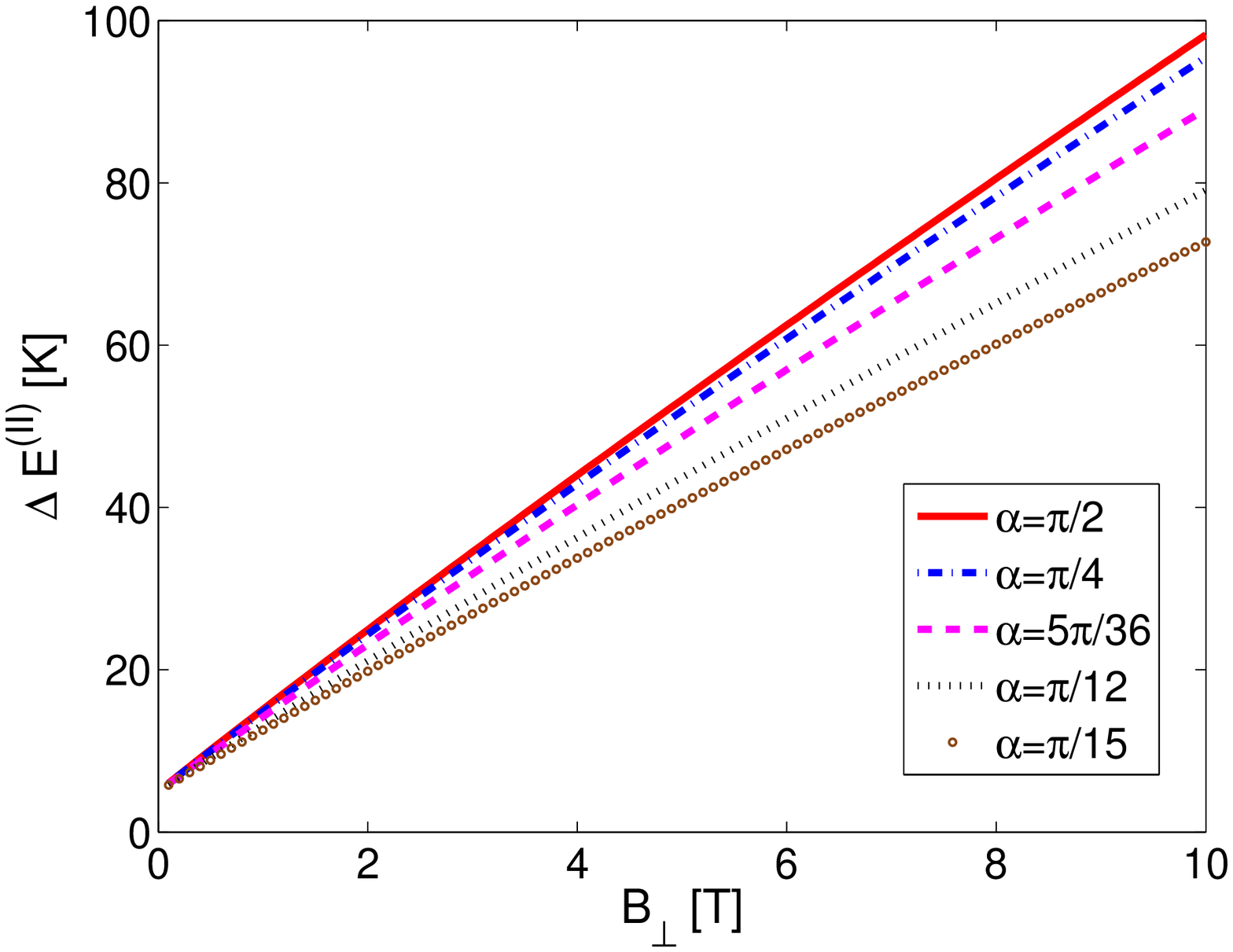}
}
\caption{The dependence of the energy gaps of solutions I (left panel)
and II (right panel) on the field $B_{\perp}$ for different angles. The parameter
$\tilde{\Delta}_{0}=5 \mbox{K}$ for both solutions I,II.}
\label{fig4}
\end{figure}

\subsection{Comparison with experiment}
\label{4b}

The first experiments in bilayer graphene in a magnetic field
\cite{Nov,Hen} revealed quantum Hall states with the
filling factor $\nu= \pm 4n$, $n= 1, 2...$ predicted in the
framework of the one electron problem in Ref. \onlinecite{McC}.
No traces of lifting the eightfold degeneracy of the
LLL and the fourfold degeneracy of higher LLs were observed.

Recent
experiments in bilayer graphene \cite{FMY,Zh} showed the generation of
energy gaps in a magnetic field resulting in
complete lifting the eightfold degeneracy in the LLL,
which leads to new quantum Hall states with filling
factors $\nu=0,\pm1,\pm2,\pm3$. While in Ref.
\onlinecite{FMY} suspended
bilayer graphene was used, bilayer graphene samples deposited on
Si${\mbox O_2}$/Si substrates were used
in Ref. \onlinecite{Zh}. Because suspended bilayer
graphene is much cleaner than that on a substrate, the new quantum Hall
states in the former start to develop at essentially smaller magnetic fields
than
in the latter. Also, the energy gaps corresponding to these states are
essentially
larger in suspended samples than in those on substrates. Both these experiments
clearly showed that the $\nu = 0$ state is an
insulating one.

Since in this paper the dynamics of the $\nu = 0$ state in
clean bilayer graphene is analyzed,
it would be appropriate to compare our results with those in suspended
graphene in more detail.
The central results concerning the $\nu = 0$ state in
Ref. \onlinecite{FMY} are: a) the observation of an extremely large
magnetoresistance in the $\nu=0$ state due to the energy gap $\Delta E$, which
scales linearly with a magnetic field $B$, $\Delta E \sim 3.5-10.5
B_{\perp}[{\mbox T}]{\mbox K}$ at least for $B_{\perp} \lesssim 10{\mbox T}$, and
b) at fixed $B_{\perp}$, an increase in the parallel component of the field
reduces the magnetoresistance at least for $B_{\perp} \lesssim 6$T.
This can be interpreted as reducing
the energy gap $\Delta E$ with increasing $B_{\parallel}$.

As to result a), the agreement of the
expressions for both the gaps
$\Delta E^{I}$ and
$\Delta E^{II}$ derived
in Sec. \ref{4a} with the gap $\Delta E$ observed in Ref. \onlinecite{FMY} is
satisfactory. Concerning the result b), it suggests that the
longitudinal magnetic field suppresses the energy gap. This fact excludes
the ferromagnetic phase as a candidate for the description of the clean
bilayer graphene at least for $B_{\perp} \leq 6$T (see left panel in
Fig. \ref{fig4}).
On the other hand, the
solution II, describing the layer asymmetric phase, is a viable candidate
for this role (see right panel in Fig. \ref{fig4}).

This conclusion together
with the phase diagram in Fig. \ref{fig1} suggest the following picture.
At $B_{\perp} < B_{cr} \sim 10$T, the layer asymmetric
phase (solution II) is realized. At
$B_{\perp} = B_{cr}$, a phase transition to the ferromagnetic phase (solution I)
takes place.
Because these solutions coexist at $B_{\perp} < B_{cr}$, one should expect that it
is a first order phase transition.
Taking literally the relation
$B_{cr}[{\mbox T}] \simeq 0.4\tilde{\Delta}_0$[K] derived from
the phase diagram in Fig. \ref{fig1} in Sec. \ref{4a},
we find that $B_{cr} \sim 10$T corresponds to $\tilde{\Delta}_0 \sim 25$K.
However,
because the existence of relevant dynamical contributions
beyond the random phase approximation is quite possible, one should consider this relation just as
a qualitative estimate.

\section{Conclusion}
\label{5}

The dynamics of bilayer graphene in a magnetic field $B \lesssim
B_{thr}$ is characterized by a very strong screening of the Coulomb
interaction that relates to the presence of a large mass $m$ in the
nonrelativistic-like dispersion relation for quasiparticles. The
functional dependence of the gap on $B$ derived in Sec. \ref{4a}
agrees with that obtained very recently in experiment in Ref.
\onlinecite{FMY}. The existence of the first order phase transition
between the layer asymmetric phase and the ferromagnetic one in the
$(\tilde{\Delta}_0,B)$ plane is predicted.

There are still many open issues in this dynamics. In particular:

 a) It would be important to include the chemical potential
$\mu_0$ in the present analysis in order to describe the higher,
$\nu = 1, 2,$ and 3, LLL plateaus \cite{FMY,Zh}.

 b) The present ansatz with the sixteen order parameters is the
minimal one for describing the breakdown of the
$U^{(K)}(2)_S \times U^{(K^{\prime})}(2)_S\times
Z_{2V}^{(+)}\times Z_{2V}^{(-)}$ symmetry in bilayer graphene.
It could be extended in order to look for other
solutions of the gap equation. A natural extension would be to
include order parameters that mix the $n=0$ and $n=1$ LLL
states.

 c) Although in Sec. \ref{3b} we presented arguments showing that the
static limit for the polarization function is at least reasonable,
it would be important to check this conclusion directly by analyzing
the gap equation with a non-static polarization function.

 d) It would be interesting to describe explicitly the dynamics around
the threshold value $B_{thr}$, when the crossover between the
regimes with the nonrelativistic-like scaling $\Delta E\sim |eB|$
and the relativistic-like one $\Delta E \sim \sqrt{|eB|}$ should
take place.

 We are planning to consider these issues elsewhere.

\begin{acknowledgments}
We  thank Junji Jia and S.G. Sharapov for fruitful discussions.
The work of E.V.G and V.P.G. was supported partially by the
SCOPES grant No. IZ73Z0\verb|_|128026 of the Swiss NSF, the grant SIMTECH No. 246937
of the European FP7 program, the grant RFFR-DFFD No. F28.2/083, and by the Program
of Fundamental Research of the Physics and Astronomy Division of the NAS of Ukraine.
The work of V.A.M. was supported by the Natural Sciences and Engineering
Research Council of Canada.

\end{acknowledgments}

\appendix
\section{Polarization operator of bilayer graphene in a magnetic field}
\label{A}

The polarization function $\Pi_{ij}$ describes electron densities
correlations on the layers $i$ and $j$:
\be
\delta(\omega+\omega^{\prime})\delta(\mathbf{k}+\mathbf{k}^{\prime})\Pi_{ij}(\omega,\mathbf{k})=
-i<0|\rho_i(\omega,\mathbf{k}^{\prime})\rho_j(\omega^{\prime},\mathbf{k}^{\prime})|0>\,.
\label{polarization-functions}
\ee
There are two independent polarization functions,
$\Pi_{11}=\Pi_{22}$ and $\Pi_{12}=\Pi_{21}$. Taking into account the polarization
effects, the bare interactions  transform into
\ba
\hat{V}_{eff}=\hat{V}\cdot\frac{1}{1+\hat{V}\cdot\hat{\Pi}}=
\left(\begin{array}{cc}\tilde{V}_{eff}(k)&\tilde{V}_{12\,eff}(k)
\\ \tilde{V}_{12\,eff}(k)& \tilde{V}_{eff}(k)\end{array}\right),\quad \hat{V}=\left(\begin{array}{cc}
\tilde{V}(k)&\tilde{V}_{12}(k)
\\ \tilde{V}_{12}(k)& \tilde{V}(k)\end{array}\right),\quad \hat{\Pi}=\left(\begin{array}{cc}
\Pi_{11}(k)&\Pi_{12}(k)
\\ \Pi_{12}(k)& \Pi_{11}(k)\end{array}\right),
\ea
with
\ba
&&\tilde{V}_{eff}(\omega,k)=\frac{2\pi e^2}{\kappa}\,\frac{k+\frac{2\pi
e^2}{\kappa}\Pi_{11}(1-e^{-2kd})} { \left[k+\frac{2\pi e^2}{\kappa}(\Pi_{11}+\Pi_{12})
(1+e^{-kd})\right]\left[k+\frac{2\pi e^2}{\kappa}(\Pi_{11}-\Pi_{12})
(1-e^{-kd})\right]}\,, \label{interaction-D}\\
&& \tilde{V}_{12\,eff}(\omega,k)=\frac{2\pi
e^2}{\kappa}\,\frac{ke^{-kd}-\frac{2\pi
e^2}{\kappa}\Pi_{12}(1-e^{-2kd})} {\left[k+\frac{2\pi e^2}{\kappa}(\Pi_{11}+\Pi_{12})
(1+e^{-kd})\right]\left[k+\frac{2\pi e^2}{\kappa}(\Pi_{11}-\Pi_{12})
(1-e^{-kd})\right]}\,, \label{interaction-ND}
\ea
where $k=|\mathbf{k}|$, and since $\Pi_{11}$ and $\Pi_{12}$ depend on
$\omega$, the effective interactions
$\tilde{V}_{eff}$ and $\tilde{V}_{12\,eff}$ depend on it too.

Neglecting the dependence on $d$ (i.e., taking $d=0$),
we obtain:
\be
\tilde{V}_{eff}(\omega,k)=\tilde{V}_{12\,eff}(\omega,k)=\frac{2\pi e^2}{\kappa}\,
\frac{1}{k+\frac{4\pi e^2}{\kappa}\Pi(\omega,k^{2})},
\label{coulomb-interaction}
\ee
where $\Pi(\omega,k^{2})\equiv \Pi_{11}(\omega,\mathbf{k})+\Pi_{12}(\omega,\mathbf{k})$
is the polarization function in a magnetic field. On the other hand,
\be
\tilde{V}_{IL}(\omega,{k})=\tilde{V}_{12\,eff}(\omega,k)-\tilde{V}_{eff}(\omega,k)=
-\frac{2\pi e^{2}}{\kappa}\frac{1-e^{-k d}}{k+\frac{2\pi e^{2}}{\kappa}
(\Pi_{11}(\omega,k)-\Pi_{12}(\omega,k))(1-e^{-k d})},
\ee
therefore, since the interlayer term
$\tilde{V}_{IL}(\omega, k)$ appears in gap equations (\ref{gap-equation-g-2}) and
(\ref{gap-equation-g-3}) only at $\omega = k = 0$,
we find that
\be
\tilde{V}_{IL}(\omega=0,{k}=0)=-\frac{2\pi e^{2}d}{\kappa_{eff}},\quad
\kappa_{eff}=\kappa+2\pi e^{2}d(\Pi_{11}(0)-\Pi_{12}(0)).
\label{interlayer-interaction}
\ee
By definition, the polarization functions $\Pi_{11}$ and $\Pi_{12}$ are defined as
\be
\Pi_{11}(\omega,\mathbf{p})=i\int \frac{d\omega^{\prime}
d^2k}{(2\pi)^3}\,\mbox{tr}\,\left[\,P_1\,\tilde{S}(\omega^{\prime},\mathbf{k})\,P_1\,
\tilde{S}(\omega+\omega^{\prime},\mathbf{p}+\mathbf{k})\,\right]\,,
\label{p11}
\ee
\be
\Pi_{12}(\omega,\mathbf{p})=i\int \frac{d\omega^{\prime}
d^2k}{(2\pi)^3}\,\mbox{tr}\,\left[\,P_1\,\tilde{S}(\omega^{\prime},\mathbf{k})\,P_2\,
\tilde{S}(\omega+\omega^{\prime},\mathbf{p}+\mathbf{k})\,\right]\,,
\label{p12}
\ee
{where $P_1=({1+\xi\tau_3})/{2}$ and $P_2=({1-\xi\tau_3})/{2}$ are projectors
on layers 1 and 2, respectively, the trace includes the summation both over the valley
index $\xi$ and spin, and $\tilde{S}(\omega,\mathbf{k})$ is the Fourier transform of the translation
invariant part of the free fermion propagator (\ref{free-propagator}) in a magnetic field.}

We are interested in calculating
the polarization function $\Pi(\omega,k^{2})$
in the random phase approximation at the neutral point ($\mu_0 = 0$).  Its
expression in configuration space is
\ba
\Pi(\omega,\mathbf{r})=i\int\frac{d\omega^{\prime}}{2\pi}{\rm tr}\left[P_{1}\tilde{S}
(\omega^{\prime},\mathbf{r})\tilde{S}(\omega+\omega^{\prime},-\mathbf{r})\right],
\ea
where a small Zeeman term in the fermion propagator will be ignored.
Then
\ba
\Pi(\omega,\mathbf{p})&=&\int d^{2}r\,e^{-i\mathbf{p}\mathbf{r}}\Pi(\omega,\mathbf{r})=
\frac{2 i}{(2\pi l^{2})^{2}}\sum\limits_{n,m=0}^{\infty}
\int\frac{d\omega^{\prime}}{2\pi}\frac{1}{[{\omega^{\prime}}^{2}-E_{n}^{2}+i0]
[(\omega+\omega^{\prime})^{2}-E_{m}^{2}+i0]}\nonumber\\
&\times&\int d^{2}r\,e^{-\mathbf{r}^{2}/2l^{2}-i\mathbf{p}\mathbf{r}}\left\{
(\omega^{\prime}+\tilde{\Delta}_{0})(\omega+\omega^{\prime}+\tilde{\Delta}_{0})
\left[L_{n}(r^{2}/2l^{2})L_{m}(r^{2}/2l^{2})+L_{n-2}(r^{2}/2l^{2})L_{m-2}(r^{2}/2l^{2})
\right]\right.\nonumber\\
&+&\left. \frac{2\hbar^{4}r^{4}}{(2m l^{4})^{2}}L^{2}_{n-2}(r^{2}/2l^{2})
L^{2}_{m-2}(r^{2}/2l^{2})\right\}.
\ea
Integrating over the angle and making the change of the variable $r^{2}=2l^{2}t$,
we get
\ba
\Pi(\omega,\mathbf{p})&=&\frac{ i}{\pi l^{2}}\sum\limits_{n,m=0}^{\infty}\int\frac{d\omega^{\prime}}{2\pi}
\frac{1}{[{\omega^{\prime}}^{2}-E_{n}^{2}+i0][(\omega+\omega^{\prime})^{2}-E_{m}^{2}+i0]}
\int\limits_{0}^{\infty}d t\, e^{-t}J_{0}(\sqrt{2p^{2}l^{2}t})\nonumber\\
&\times&\left\{(\omega^{\prime}+\tilde{\Delta}_{0})(\omega+\omega^{\prime}+\tilde{\Delta}_{0})
\left[L_{n}(t)L_{m}(t)+L_{n-2}(t)L_{m-2}(t)\right]+2\omega_{c}^{2}t^{2}L^{2}_{n-2}(t)L^{2}_{m-2}(t)\right\},
\label{A10}
\ea
where $J_{\nu}$ is a Bessel function and
$E_n = \sqrt{\hbar^2\omega_{c}^{2}n(n-1)+\tilde{\Delta}_{0}^{2}}$ (compare with
Eq. (\ref{free-propagator})).

In order to evaluate the $t$ integral with the first
term in the curl brackets in Eq. (\ref{A10}),
we will use the formula 7.422.2 in Ref. \onlinecite{GR}:
\be
\int\limits_{0}^{\infty}dx x^{\nu+1}\,e^{-\alpha x^{2}}J_{\nu}(b x)L_{m}^{\nu-\sigma}(\alpha x^{2})
L_{n}^{\sigma}(\alpha x^{2})=(-1)^{m+n}(2\alpha)^{-\nu-1}b^{\nu}
\exp\left(-\frac{b^{2}}{4\alpha}\right) L_{m}^{\sigma-m+n}\left(\frac{b^{2}}{4\alpha}\right)
L_{n}^{\nu-\sigma+m-n}\left(\frac{b^{2}}{4\alpha}\right).
\label{formula1}
\ee
Taking $\nu=\sigma=0, \alpha=1, b=2\sqrt{y},\,y={p^{2}l^{2}}/2$ in this
expression, we obtain
\be
\int\limits_{0}^{\infty}d t\, e^{-t}J_{0}(2\sqrt{y t})L_{n}(t)L_{m}(t)=(-1)^{m+n}
e^{-y}L_{m}^{n-m}(y)L_{n}^{m-n}(y)\equiv{(-1)^{m+n}}
e^{-y}I_{nm}(y)
\ee
with
\be
I_{nm}(y)= L_{m}^{n-m}(y)L_{n}^{m-n}(y).
\ee
At small $y$, we find
\be
I_{nm}(y)\simeq\delta_{nm}-y\left[2n\delta_{nm}+(m+1)\delta_{n,m+1}
+(n+1)\delta_{m,n+1}\right],\quad y\to0.
\label{asymptotic-Inm}
\ee

In order to evaluate the $t$ integral with the second term
in the curl brackets in Eq. (\ref{A10}),
\be
\int\limits_{0}^{\infty}d t\,t^{2}\, e^{-t}J_{0}(2\sqrt{y t})L^{2}_{n}(t)L^{2}_{m}(t)
\equiv{(-1)^{m+n}}e^{-y}I^{(2)}_{nm}(y),
\ee
{we set $\nu=0,\sigma=2, b=2\sqrt{y}$ in Eq.(\ref{formula1}),}
\ba
\int\limits_{0}^{\infty}dx\,x\,e^{-x^{2}}J_{0}(2x\sqrt{y})L^{-2}_{m}(x^{2})L^{2}_{n}(x^{2})
=\frac{(-1)^{m+n}}{2}e^{-y}L^{2-m+n}_{m}(y)L^{-2+m-n}_{n}(y),
\ea
and use the following identity for the Laguerre polynomials on the left hand side of
this equation,
\be
L_{l}^{k}(x)=(-x)^{-k}\frac{(l+k)!}{l!}L^{-k}_{l+k}(x),\quad l\ge0,\, k+l\ge0.
\label{Laguerre-identity}
\ee
Then we arrive at
\ba
\int\limits_{0}^{\infty}dt\,t^{2}\,e^{-t}J_{0}(2\sqrt{y t})L^{2}_{m-2}(t)L^{2}_{n}(t)
=(-1)^{m+n}\frac{m!}{(m-2)!}\,e^{-y}L^{2+n-m}_{m}(y)L^{-2+m-n}_{n}(y),
\ea
and, therefore,
\be
I^{(2)}_{nm}(y)=(m+1)(m+2)L^{n-m}_{m+2}(y)L^{m-n}_{n}(y).
\label{I2nm}
\ee
Although the symmetry of $I^{(2)}_{nm}(y)$ under the interchange $n\leftrightarrow m$
is not explicit, it can be checked by using the identity (\ref{Laguerre-identity}).
At small $y$, we get the following expansion for $I^{2}_{nm}(y)$:
\ba
I^{(2)}_{nm}(y)&\simeq&(n+1)(n+2)\delta_{nm}-y\left[2(n+1)^{2}(n+2)\delta_{nm}
\right.\nonumber\\
&+&\left.\delta_{n,m+1}n(n+1)(n+2)+\delta_{m,n+1}m(m+1)(m+2)\right],\quad y\to0.
\label{asymptotic-I2nm}
\ea
Therefore the polarization function (\ref{A10}) takes the following form:
\ba
\Pi(\omega,\mathbf{p})&=&\frac{2 i\,e^{-y}}{(2\pi l)^{2}}\sum\limits_{n,m=0}^{\infty}(-1)^{m+n}
\int\limits_{-\infty}^{\infty}\frac{d\omega^{\prime}}{[{\omega^{\prime}}^{2}-E_{n}^{2}+i0]
[(\omega+\omega^{\prime})^{2}-E_{m}^{2}+i0]}\nonumber\\
&\times&\left\{(\omega^{\prime}+\tilde{\Delta}_{0})(\omega+\omega^{\prime}+\tilde{\Delta}_{0})
\left[I_{nm}(y)+I_{n-2,m-2}(y)\right]+2\omega_{c}^{2}I^{(2)}_{n-2,m-2}(y)\right\}.
\label{Pi-energy-momentum}
\ea
After integrating over $\omega^{\prime}$ in this expression, we obtain
\ba
\Pi(\omega,\mathbf{p})=\frac{ \,e^{-y}}{2\pi l^{2}}\sum\limits_{n,m=0}^{\infty}
\frac{(-1)^{m+n}(E_{n}+E_{m})}{(E_{n}+E_{m})^{2}-\omega^{2}}\left[\left(1-
\frac{\tilde{\Delta}_{0}^{2}}{E_{n}E_{m}}\right)[I_{nm}(y)
+I_{n-2,m-2}(y)]-\frac{2\omega_{c}^{2}}{E_{n}E_{m}}I^{(2)}_{n-2,m-2}(y)\right]
\label{Pi:frequency}
\ea
with $I_{nm},I^{2}_{nm}\equiv0$ for $n<0$ or $m<0$. It is noticeable that the
contribution of the LLL 
{(with $n,m = 0, 1$ and $E_0 = E_1 = \tilde{\Delta}_{0}$) 
in the polarization function is identically zero.}

{Let us discuss the properties of the static polarization used in the
main text in more detail.}
For the static polarization function
$\Pi(\omega=0,\mathbf{p})$ in balanced bilayer graphene,
$\tilde{\Delta}_{0}=0$, we get
\ba
\Pi(0,\mathbf{p})&=&\frac{m}{2\pi\hbar^{2}}\,e^{-y}\left\{\sum\limits_{n,m=2}^{\infty}
\frac{(-1)^{n+m}}{M_{n}M_{m}(M_{n}+M_{m})}
\left[M_{n}M_{m}\left(I_{n-2,m-2}(y)+I_{nm}(y)\right)-2
I^{(2)}_{n-2,m-2}(y)\right]\right.\nonumber\\
&+&\left.2\sum\limits_{n=2}^{\infty}\frac{(-1)^{n}}{M_{n}}
\left[I_{0n}(y)-I_{1,n}(y)\right]\right\}\equiv \frac{m}{\hbar^{2}}\tilde{\Pi}(y),
\label{Pi:static}
\ea
where $M_{n}=\sqrt{n(n-1)}$.
Note that the quasiparticle mass $m$ appears as an overall factor only and
does not enter the function $\tilde{\Pi}(y)$. We checked that the double
sum is convergent and in numerical calculation we took the upper limits in the sum around
$n_{max}, m_{max}=250$, this is enough to calculate $\tilde{\Pi}(y)$ up to values $y=12$
as it is shown in Fig. \ref{fig1}.

Using Eqs.(\ref{asymptotic-Inm}), (\ref{asymptotic-I2nm}), and
(\ref{Pi:static}),
we find
the asymptotics of the static polarization function at $y\to0$:
\be
\tilde{\Pi}(y)\simeq\frac{2y}{\pi}\left[\sum\limits_{n=2}^{\infty}\frac{1}{\sqrt{n}
(\sqrt{n+1}+\sqrt{n-1})(n+\sqrt{n^{2}-1})}+\frac{1}{\sqrt{2}}\right]\approx0.55y.
\ee
The main contribution in this expression (around $80\%$)
comes from the transitions between the LLL
and the first higher LL with $n = 2$ (the term $1/\sqrt{2}$ in brackets).
With increasing $y$, the number of higher LLs providing
relevant contributions in the polarization function grows.

In a similar way, one can find the expressions
for the two independent polarization functions $\Pi_{11}(0,\mathbf{p})$
and $\Pi_{12}(0,\mathbf{p})$:
\ba
\Pi_{11}(y)&=&\frac{m\,e^{-y}}{2\pi\hbar^{2}}\left\{\sum\limits_{n,m=2}^{\infty}
\frac{(-1)^{n+m}}{M_{n}M_{m}(M_{n}+M_{m})}
\left[M_{n}M_{m}\left(I_{n-2,m-2}(y)+I_{nm}(y)\right)\right]
+2\sum\limits_{n=2}^{\infty}\frac{(-1)^{n}}{M_{n}}
\left[I_{0n}(y)-I_{1,n}(y)\right]\right\},\nonumber\\
\Pi_{12}(y)&=&-\frac{m\,e^{-y}}{\pi\hbar^{2}}\sum\limits_{n,m=2}^{\infty}
\frac{(-1)^{n+m}}{M_{n}M_{m}(M_{n}+M_{m})}I^{(2)}_{n-2,m-2}(y).
\ea
At zero momentum, $y=0$, we have
\be
\Pi_{11}(y=0)=-\Pi_{12}(y=0)=\frac{m}{2\pi\hbar^{2}}\sum\limits_{n=2}^{n_{max}}\frac{1}{\sqrt{n(n-1)}}.
\ee
As is seen, the quantities $\Pi_{11},\Pi_{12}$ are logarithmically divergent separately,
and we introduced cutoff $n_{max}$.
The physical origin of this cutoff is the following. For high energy modes,
monolayer like dynamics takes place, whose contribution to $\Pi_{11}(y=0)$ and
$\Pi_{12}(y=0)$ is strongly suppressed (recall that in monolayer graphene $\Pi(0) = 0$,
see for example Ref. \onlinecite{Gorb}). Therefore the cutoff $n_{max}$ can roughly
be estimated from the condition of the applicability of
the low energy effective model \cite{McC}:
$\hbar\omega_{c}\sqrt{n(n-1)}<\gamma_{1}/4$, which gives $n_{max} \simeq
(\gamma_{1}l/2\sqrt{2}\hbar v_{F})^{2}\approx 45/B[{\mbox T}]$ (here
the values {m}, $\gamma_{1}$, and $v_{F}$ are taken from Ref. \onlinecite{McC}).

For the effective dielectric constant $\kappa_{eff}$ in Eq.(\ref{interlayer-interaction})
we thus get:
\be
\kappa_{eff}=\kappa+\frac{2m e^{2}d}{\hbar^{2}}\sum\limits_{n=2}^{n_{max}}\frac{1}{\sqrt{n(n-1)}}
=\kappa+0.68\sum\limits_{n=2}^{n_{max}}\frac{1}{\sqrt{n(n-1)}},
\ee
where the value $m\approx0.054 m_{e}$ was used. With $n_{max}\approx 45/B[{\mbox T}]$, we
find that for the range of fields
from $0.25$T to 10T the cutoff $n_{max}$ varies in the interval $6\div180$,
and therefore the quantity
$\kappa_{eff}-\kappa$ varies in the interval $1\div4$.


\begin{thebibliography}{99}
\bibitem{footnote}
One can show that the presence of a maximum in
the function $4\pi \tilde{\Pi}(y)$ in Fig. \ref{fig1}
follows from the equality of
the polarization charge density $n(r)$ in a magnetic field
$B$ and that at $B=0$ as $r \to 0$.
\end{thebibliography}
\end{document}